\input harvmac
\input epsf
\noblackbox
\def\npb#1#2#3{{\it Nucl.\ Phys.} {\bf B#1} (#2) #3}
\def\plb#1#2#3{{\it Phys.\ Lett.} {\bf B#1} (#2) #3}

\def\prd#1#2#3{{\it Phys.\ Rev.} {\bf D#1} (#2) #3}
\def\rmp#1#2#3{{\it Rev.\ Mod.\ Phys.} {\bf #1} (#2) #3}
\def\mpla#1#2#3{{\it Mod.\ Phys.\ Lett.} {\bf A#1} (#2) #3}

\def\jmp#1#2#3{{\it J. Math.\ Phys.} {\bf #1} (#2) #3}
\def\cqg#1#2#3{{\it Class.\ Q. Grav.} {\bf #1} (#2) #3}

\def\jhep#1#2#3{{\it JHEP\/} {\bf #1} (#2) #3}

\newcount\figno
\figno=0
\def\fig#1#2#3{
\par\begingroup\parindent=0pt\leftskip=1cm\rightskip=1cm\parindent=0pt
\baselineskip=11pt
\global\advance\figno by 1
\midinsert
\epsfxsize=#3
\centerline{\epsfbox{#2}}
\vskip 12pt
{\bf Fig.\ \the\figno: } #1\par
\endinsert\endgroup\par
}
\def\figlabel#1{\xdef#1{\the\figno}}
\def\encadremath#1{\vbox{\hrule\hbox{\vrule\kern8pt\vbox{\kern8pt
\hbox{$\displaystyle #1$}\kern8pt}
\kern8pt\vrule}\hrule}}
\def\berkeley{\centerline{\it Berkeley Center for Theoretical Physics and 
Department of Physics}
\centerline{\it University of California, Berkeley, CA 94720-7300}
\centerline{\it and}
\centerline{\it Theoretical Physics Group, Lawrence Berkeley National 
Laboratory}
\centerline{\it Berkeley, CA 94720-8162, USA}}

\def\arcsinh{{\rm arcsinh}\,}

\def\godel{G\"odel}
\def\endli{\hfill\break}
\def\frac#1#2{{#1 \over #2}}

\def\p{\partial}
\def\semi{\subset\kern-1em\times\;}
\def\CA{{\cal A}}                   
                   \def\CG{{\cal G}}
\def\CH{{\cal H}}                   
                   \def\CL{{\cal L}}

\def\CR{{\cal R}}

\def\R{{\bf R}}                     \def\S{{\bf S}}
                     
%

%
\Title{\vbox{\baselineskip12pt
\hbox{hep-th/0212087}
\hbox{LBNL-51811}
\hbox{UCB-PTH-02/57}}}
{\vbox{\centerline{Holographic Protection of Chronology}
\bigskip
\centerline{in Universes of the G\"odel Type}}}
\bigskip
\centerline{Edward K. Boyda, Surya Ganguli, Petr Ho\v rava and Uday 
Varadarajan}
\bigskip
\baselineskip14pt
\berkeley
\centerline{\tt ekboyda, sganguli, horava, udayv@socrates.berkeley.edu}
\medskip\bigskip\medskip
\noindent
We analyze the structure of supersymmetric G\"odel-like cosmological solutions 
of string theory.  Just as the original four-dimensional G\"odel universe, 
these solutions represent rotating, topologically trivial cosmologies with a 
homogeneous metric and closed timelike curves.  First we focus on 
``phenomenological'' aspects of holography, and identify the preferred 
holographic screens associated with inertial comoving observers in \godel\ 
universes.  We find that holography can serve as a chronology protection 
agency: The closed timelike curves are either hidden behind the holographic 
screen, or broken by it into causal pieces.  In fact, holography 
in \godel\ universes has many features in common with de~Sitter space, 
suggesting that \godel\ universes could represent a supersymmetric laboratory 
for addressing the conceptual puzzles of de~Sitter holography.  Then we 
initiate the investigation of ``microscopic'' aspects of holography of \godel\ 
universes in string theory.  We show that \godel\ universes are T-dual to 
pp-waves, and use this fact to generate new \godel-like solutions of 
string and M-theory by T-dualizing known supersymmetric pp-wave solutions.  
\Date{December 2002}
\nref\holog{G.~'t~Hooft, ``Dimensional Reduction in Quantum Gravity,'' 
gr-qc/9310026\endli
L.~Susskind, ``The World as a Hologram,'' \jmp{36}{1995}{6377},\hfill\break
hep-th/9409089.}
\nref\tomds{T. Banks, ``Cosmological Breaking of Supersymmetry or 
Little Lambda Goes Back to the Future II,'' hep-th/0007146\hfill\break
see also T. Banks and W. Fischler, ``M-Theory Observables for 
Cosmological Space-Times,'' hep-th/0102077\hfill\break
W. Fischler, unpublished (2000)\hfill\break
W. Fischler, ``Taking de~Sitter Seriously,'' talk given at 
{\it Role of Scaling Laws in Physics and Biology (Celebrating the 60th 
Birthday of Geoffrey West)}, Santa Fe, December 2000 (unpublished).}
\nref\lds{L. Susskind, ``Twenty Years of Debate with Stephen,'' 
hep-th/0204027.}
\nref\dls{L. Dyson, J. Lindesay and L. Susskind, ``Is There Really a 
de~Sitter/CFT Duality,'' \jhep{0208}{2002}{045}.}
\nref\raphaelds{R. Bousso, ``Positive Vacuum Energy and the N-Bound,'' 
\jhep{0011}{2000}{038}, hep-th/0010252.}
\nref\petrds{P. Ho\v rava, talk at Strings 2001 at Mumbai, India (January 
2001)\endli
V. Balasubramanian, P. Ho\v rava and D. Minic, ``Deconstructing de~Sitter,'' 
\jhep{0105}{2001}{043}, hep-th/0103171.}
\nref\edds{E. Witten, talk at Strings 2001 in Mumbai, India (January 2001)
\hfill\break
E. Witten, ``Quantum Gravity in de~Sitter Space,'' hep-th/0106109.}
\nref\andyds{A. Strominger, ``The dS/CFT Correspondence,'' 
\jhep{0110}{2001}{034}, hep-th/0106113\hfill\break
see also M. Spradlin, A. Strominger and A. Volovich, ``Les Houches Lectures 
on de~Sitter Space,'' hep-th/0110007.}
\nref\gdl{K. G\"odel, ``An Example of a New Type of Cosmological Solutions 
of Einstein's Field Equations of Gravitation,'' \rmp{21}{1949}{447}.}
\nref\jerome{J.P. Gauntlett, J.B Gutowski, C.M. Hull, S. Pakis and H.S. Reall, 
``All Supersymmetric Solutions of Minimal Supergravity in Five Dimensions,'' 
hep-th/0209114.}
\nref\tseytlin{A.A. Tseytlin, ``Extreme Dyonic Black Holes in String Theory,'' 
\mpla{11}{1996}{689}, hep-th/9601177.}
\nref\gibher{G.W. Gibbons and C.A.R. Herdeiro, ``Supersymmetric Rotating 
Black Holes and Causality Violation,'' \cqg{16}{1999}{3619}, 
hep-th/9906098\hfill\break
C.A.R. Herdeiro, ``Special Properties of Five Dimensional 
BPS Rotating Black Holes,'' \npb{582}{2000}{363}, hep-th/0003063.}
\nref\visser{M. Visser, {\it Lorentzian Wormholes: From Einstein to 
Hawking\/} (AIP Press, 1995).} 
\nref\carter{B. Carter, ``Global Structure of the Kerr Family of Gravitational 
Fields,'' {\it Phys.\ Rev.} {\bf 174} (1968) 1559.}
\nref\raphael{R. Bousso, ``Holography in General Space-times,'' 
\jhep{9906}{1999}{028}, hep-th/9906022.}
\nref\holorev{D. Bigatti and L. Susskind, ``TASI Lectures on the Holographic 
Principle,'' hep-th/0002044\hfill\break
R. Bousso, ``The Holographic Principle,'' \rmp{74}{2002}{825}, 
hep-th/0203101.}
\nref\elsewhere{E.K. Boyda, S. Ganguli, P. Ho\v rava and U. Varadarajan, 
work in progress.}
\nref\hawkell{S.W. Hawking and G.F.R. Ellis, {\it The Large Scale Structure of 
Space-Time\/} (Cambridge University Press, Cambridge, 1973).}
\nref\abs{R. Adler, M. Bazin and M. Schiffer, {\it Introduction to 
General Relativity\/} (Mc Graw-Hill, 1965).}
\nref\chandra{S. Chandrasekhar and J.P. Wright, ``The Geodesics in \godel's 
Universe,'' {\it Proc.\ Nat.\ Acad.\ Sci.} {\bf 47} (1961) 341.}
\nref\chrono{S.W. Hawking, ``The Chronology Protection Conjecture,'' 
\prd{46}{1992}{603}.}
\nref\hommet{A.K. Raychaudhuri and S.N.G. Thakurta, ``Homogeneous Space-times 
of the \godel\ Type,'' \prd{22}{1980}{802}\endli
M.J. Rebou\c cas and J. Tiomno, ``Homogeneity of Riemannian Space-times of 
\godel\ Type,'' \prd{28}{1982}{1251}.}
\nref\squash{M. Rooman and Ph.\ Spindel, ``G\"odel Metric as a Squashed 
Anti-de~Sitter Geometry,'' \cqg{15}{1998}{3241}, gr-qc/9804027.}
\nref\somr{M.M. Som and A.K. Raychaudhuri, ``Cylindrically Symmetric Charged 
Dust Distributions in Rigid Rotation in General Relativity,'' 
{\it Proc.\ Roy.\ Soc.\ London\/} {\bf A304} (1968) 81.}
\nref\stockum{W.J. van Stockum, ``Gravitational Field of a Distribution of 
Particles Rotating About an Axis of Symmetry,'' {\it Proc.\ Roy.\ S. Edin.} 
{\bf 57} (1937) 135\hfill\break
F.J. Tipler, ``Rotating Cylinders and the Possibility of Global 
Causality Violations,'' \prd{9}{1974}{2203}\hfill\break
P. Klep\'a\v c and J. Horsk\'y, ``''Charged Perfect Fluid and Scalar Field 
Coupled to Gravity,'' {\it Czech.\ J. Phys.}  {\bf 51} (2001) 1177, 
gr-qc/0109043.}
\nref\townsend{J.P. Gauntlett, R.C. Myers and P.K. Townsend, ``Black Holes 
of $D=5$ Supergravity,'' \cqg{16}{1999}{1}, hep-th/9810204\hfill\break
P.K. Townsend, ``Surprises with Angular Momentum,'' hep-th/0211008.}
\nref\recurr{L. Dyson, M. Kleban and L. Susskind, ``Disturbing Implications 
of a Cosmological Constant,'' hep-th/0208013\hfill\break
T. Banks, W. Fischler and S. Paban, ``Recurring Nightmares? : 
Measurement Theory in de~Sitter Space,'' hep-th/0210160.}
\nref\cvetic{M. Cveti\v c, H. L\"u and C.N. Pope, ``Penrose Limits, PP-Waves 
amd Deformed M2-Branes,'' hep-th/0203082.}
\nref\tpp{R. G\"uven, ``Plane Wave Limits and T-Duality,'' 
\plb{482}{2000}{255}, hep-th/0005061
\hfill\break
J. Michelson, ``(Twisted) Toroidal Compactification of pp-Waves,'' 
\prd{66}{2002}{066002}, hep-th/0203140.}
\nref\iibpp{M. Blau, J. Figueroa-O'Farrill, C. Hull and G. Papadopoulos, 
``A New Maximally Supersymmetric Background of IIB Superstring Theory,'' 
\jhep{0201}{2002}{047}, hep-th/0110242.}
\nref\mtw{C.W. Misner, K.S. Thorne and J.A. Wheeler, {\it Gravitation} 
(Freeman, 1973).}
\nref\lastweek{C.A.R. Herdeiro, ``Spinning Deformations of the D1-D5 System 
and a Geometric Resolution of Closed Timelike Curves,'' hep-th/0212002.}
\newsec{Introduction}

Many long-standing conceptual questions of quantum gravity, and even of 
classical general relativity, are finding their answers in string theory.  
Among the most notable examples are various classes of supersymmetric 
timelike singularities, or the microscopic explanation of Bekenstein-Hawking 
entropy for a class of configurations controllable by spacetime 
supersymmetry.  On the other hand, many puzzles of quantum gravity still 
remain unanswered.  In particular, the role of time in cosmological, and other 
time-dependent, solutions of string theory still defies any systematic 
understanding.  

While many crucial questions of quantum gravity are associated with high 
spacetime curvature or with cosmological horizons, some puzzles become 
apparent already in spacetimes with very mild curvature, no horizons, and 
even trivial topology.  How can the low-energy classical relativity fail 
to represent a good approximation to quantum gravity for small curvature 
and in the absence of horizons?  Arguments leading to the holographic 
principle \holog\ indicate that general relativity misrepresents the true 
degrees of freedom of quantum gravity, by obscuring the fact that they are 
secretly holographic.  In those instances where string theory has been 
successful in resolving puzzles of quantum gravity, it has done so by 
identifying the correct microscopic degrees of freedom, which frequently 
are poorly reflected by the naive (super)gravity approximation.  In this 
paper we investigate an example in which holography suggests a very 
specific dramatic modification of the degrees of freedom in quantum gravity 
already at very mild curvatures, in a homogeneous and highly supersymmetric 
cosmological background.  

Historically, microscopic holography in string theory has been relatively 
easier to understand for solutions with a ``canonical'' preferred holographic 
screen which is observer-independent, and typically located at asymptotic 
infinity.  Holography in $AdS$ spaces is a prime example of this.  On the 
other hand, cosmological backgrounds in string theory require an 
understanding of holography in more complicated environments, which may 
not exhibit canonical, observer-independent preferred screens at conformal 
infinity.  Here, the prime example is given by de~Sitter space: When viewed 
from the perspective of an inertial observer living in the static patch, 
the preferred holographic screen in de~Sitter space is most naturally placed 
at the cosmological horizon.  This leads to the fascinating idea of 
observer-dependent holographic screens, associated with a finite number of 
degrees of freedom accessible to the observer (for more details, see e.g.\ 
\refs{\tomds-\petrds}; see also \refs{\edds,\andyds} for a 
complementary point of view on de~Sitter holography that uses other preferred 
screens, not associated with an inertial observer).  

Of course, string theory promises to be a unified theory of gravity and 
quantum mechanics, but it is at present unclear how it 
manages to reconcile the general relativistic concept of time (notoriously 
difficult because of spacetime diffeomorphism invariance) with the quantum 
mechanical role of time as an evolutionary Hamiltonian parameter.  Again, 
this problem becomes somewhat trivialized in the presence of supersymmetry, 
but persists in all but the most trivial time-dependent backgrounds of string 
theory. 

In this paper, we analyze a class of supersymmetric solutions of string theory 
and M-theory, which -- at least in the classical supergravity approximation 
-- are described by geometries with no global time function.  In particular, 
we focus our attention on string theory analogs of \godel's universe.  
\godel's original solution \gdl\ is a homogeneous rotating cosmological 
solution of Einstein's equations with pressureless matter and negative 
cosmological constant, which played an important role in the conceptual 
development of general relativity.  Recently, a supersymmetric generalization 
of \godel's universe has been discussed in a remarkable paper by Gauntlett 
{\it et al.} \jerome , who classified all supersymmetric solutions of 
five-dimensional supergravity with eight supercharges, and found a maximally 
supersymmetric \godel-like solution that can be lifted to a solution of 
M-theory with twenty Killing spinors.  The existence of this solution was 
also noticed previously by Tseytlin, see Footnote~26 of \tseytlin .  It is 
worth stressing that the \godel\ universe of M-theory is time-orientable: 
There is an invariant notion of future and past lightcones, at each point 
in spacetime.  Also, there is a global time {\it coordinate\/} $t$, and in 
fact $\p/\p t$ is an everywhere time-like Killing vector (in effect, making 
supersymmetry possible).  However, $t$ is not a global time {\it function\/}:  
The surfaces of constant $t$ are not everywhere spacelike.%
\foot{See, e.g., \gibher\ for a detailed discussion of the distinction between 
a global time coordinate and a global time function.} 
Actually, the solution cannot be foliated by everywhere-spacelike surfaces at 
all -- the classical Cauchy problem is always ill-defined in this spacetime.  
It is hard to imagine how such an apparently pathological behavior of global 
time could be compatible with the conventional role of time in the Hamiltonian 
picture of quantum mechanics.  Indeed, this solution turns out to have 
classical pathologies: Just as \godel's original solution, the supersymmetric 
\godel\ metric allows closed timelike curves, seemingly suggesting 
either the possibility of time travel (cf.\ \visser) or at least grave 
causality problems.  

These classical pathologies could imply that the \godel\ solution, despite 
its high degree of supersymmetry, stays inconsistent even in full string 
or M-theory.  There are of course pathological solutions of Einstein's 
equations whose problems do not get resolved in string theory, with the 
negative-mass Schwarzschild black hole being one example.  

However, there are reasons why one might feel reluctant to discard 
this solution as manifestly unphysical, despite the sicknesses of the 
classical metric: This solution is homogeneous, its curvature 
can be kept small everywhere (in particular, there are no singularities and 
no horizons), and the solution is highly supersymmetric.  It is also 
impossible to eliminate the closed timelike curves by going to a universal 
cover -- indeed, the \godel\ solution is already topologically trivial.%
\foot{This should be contrasted with the case of solutions with 
``trivial'' (in the sense of Carter \carter) closed timelike curves, such as 
those in the flat Minkowski spacetime with time compactified on $\S^1$, where 
the closed timelike curves can be eliminated by lifting the solution to its 
universal cover.} 

We feel that any solution should be presumed consistent until proven 
otherwise, and this will be our attitude towards the \godel\ solution in 
this paper.  Our aim will be to analyze holographic properties of the 
supersymmetric \godel\ solution in string theory.  The solution is 
remarkably simple, and as we will see in Section~5, turns out to be related 
by duality to the solvable supersymmetric pp-wave backgrounds much 
studied recently.  However, before we attempt the analysis of 
``microscopic'' holography in string theory, we will first adopt a more 
``phenomenological'' approach as pioneered by Bousso \raphael\ (see 
\refs{\lds,\holorev} for reviews), and analyze the structure of preferred 
holographic screens implied by the covariant prescription \raphael\ for their 
identification in classical (super)gravity solutions.  This 
``phenomenological'' analysis  leads to valuable hints, indicating how the 
problem of closed timelike curves may be resolved in the \godel\ universe.  
Indeed, we will claim that the apparent pathologies of the semiclassical 
supergravity solution can be resolved when holography is properly taken into 
account.  Semiclassical general relativity without holography is not a good 
approximation of this solution, despite its small curvature, absence of 
horizons, and trivial spacetime topology.  

Notice also that homogeneity of the \godel\ solution makes things 
at least superficially worse: It implies that there are closed timelike curves 
through every point in spacetime.  However, these closed timelike curves are 
also in a sense (to be explained below) topologically ``large.''  Our analysis 
of the structure of holographic screens in this geometry reveals an intricate 
system of observer-dependent preferred holographic screens, which always carve 
out a causal part of spacetime, and effectively screen all the closed timelike 
curves and hide any violations of causality from the inertial observer.  
In fact, the causal structure of the part of spacetime carved out 
by the screen is precisely that of an AdS space, cut off at some finite radial 
distance.     

The preferred holographic screens in the \godel\ universe are very much like 
the screens associated with the inertial observers in the static patch of 
de~Sitter space.  First of all, they are associated with the selection of an 
observer (and therefore represent ``movable,'' non-canonical screens, not 
located at conformal infinity).  Moreover, they are compact, implying a 
finite covariant 
bound on entropy and -- in the strong version of the holographic principle -- 
a finite number of degrees of freedom associated with any inertial observer.  
Thus, the \godel\ universe should serve as a useful supersymmetric laboratory 
for addressing some of the conceptual puzzling issues of de~Sitter 
holography.  

The results of our ``phenomenological'' analysis of holography also reveal 
the importance, for cosmological spacetimes, of a local description of physics 
as associated with an observer inside the universe.  It is not sensible to 
pretend that the observer stays at asymptotic infinity, and observes only 
elements of the traditionally defined S-matrix (or some suitable analogs 
thereof).  Clearly, this only stresses the need for a conceptual framework 
defining more environmentally-friendly, ``cosmological'' observables 
as associated with cosmological observers in string theory.  

The structure of the paper is as follows.  In Section~2, we set the stage 
by reviewing and analyzing \godel's cosmological solution $\CG_3\times\R$ 
of Einstein's gravity in four space-time dimensions.  Despite its simplicity,
this solution already exhibits all the crucial issues of our argument.  We 
apply Bousso's prescription for the covariant holographic screens, and find 
screens that are observer-dependent, compact, and causality-preserving.  
In addition, we establish connection with holography in {\it AdS} spaces: 
\godel's solution can be viewed as a member of a two-parameter moduli space 
of homogeneous solutions of Einstein's equations with trivial spacetime 
topology, with $AdS_3\times\R$ also in this moduli space.  We show that under 
the corresponding deformation the observer-dependent preferred holographic 
screens of \godel's universe recede to infinity and become the canonical 
holographic boundary of $AdS_3\times\R$.  In Section~3 we move on to the 
supersymmetric \godel\ universe of M-theory, which can be written as 
$\CG_5\times\R^6$.  First we analyze the $\CG_5$ part of the geometry as 
a solution of minimal $d=5$ supergravity, study in detail the structure of 
geodesics in this solution and use it to determine the preferred holographic 
screens, and show how chronology can be protected by holography.   Then we 
extend our analysis to the full $\CG_5\times\R^6$ \godel\ geometry in 
M-theory.  Section~4 points out remarkable analogies between holography in the 
supersymmetric \godel\ universe and holography in de~Sitter space.  
In Section~5, we embark on the analysis of ``microscopic'' duality of \godel\ 
universes in string theory.  First, we compactify the M-theory solution on 
$\S^1$ to obtain a \godel\ solution of Type IIA superstring theory, and 
show that upon further $\S^1$ compactification the Type IIA \godel\ universe 
is T-dual to a supersymmetric Type IIB pp-wave, which can be obtained as the 
Penrose limit of the intersecting D3-D3 system. We point out that this 
\godel/pp-wave T-duality is a much more general phenomenon, and can be used 
to construct new \godel\ universes in string and M-theory by T-dualizing 
known pp-waves.  The relation to pp-waves is just one aspect of the remarkable 
degree of solvability of \godel\ solutions in string theory.  We intend to 
present a more detailed analysis of ``microscopic'' aspects of holography in 
the \godel\ universes of string and M-theory elsewhere \elsewhere .  
In Appendix~A we summarize some geometric properties of the supersymmetric 
\godel\ solutions.  

\newsec{Holography in \godel's Four-Dimensional Universe}

\subsec{\godel's solution}

In 1949, on the occasion of Albert Einstein's 70th birthday, Kurt \godel\ 
presented a rotating cosmological solution \gdl\ of Einstein's equations with 
negative cosmological constant and pressureless matter; this solution is 
topologically trivial and homogeneous but exhibits closed timelike curves.  
Our exposition of \godel's solution follows \refs{\gdl,\hawkell}.

The spacetime manifold of this solution has the trivial topology of $\R^4$, 
which we will cover by a global coordinate system $(\tau,x,y,z)$.  The 
metric factorizes into a direct sum of the (trivial) metric $dz^2$ on $\R$ 
and a nontrivial metric on $\R^3$,
\eqn\eegdfour{ds^2_4=ds^2_3+dz^2,}
where
\eqn\eegdthree{ds^2_3=-d\tau^2+dx^2-\half e^{4\Omega x}dy^2-2e^{2\Omega x}
d\tau\,dy.}
This class of solutions is characterized by a rotation parameter $\Omega$.  
We will refer to the manifold $\R^3$ equipped with the non-trivial part 
\eegdthree\ of \godel's metric as $\CG_3$.  Thus, in our notation, 
\godel's universe is $\CG_3\times\R$.  The metric on $\CG_3$ has a 
four-dimensional group of isometries.  The geometry exhibits dragging of 
inertial frames, associated with rotation.   The full four-dimensional 
geometry solves Einstein's equations, with the value of the cosmological 
constant and the density of pressureless matter both determined by the 
rotation parameter $\Omega$, 
\eqn\eedensity{\rho=\frac{\Omega^2}{2\pi G_N}, \qquad\Lambda=-2\Omega^2.}
Historically, this solution was instrumental in the discussion of whether 
or not classical general relativity satisfies Mach's principle (see, e.g., 
\abs , Sect.~12.4).  

While the homogeneity of \godel's universe is (almost) manifest in the 
coordinate system used in \eegdthree , the rotational symmetry of $ds^2_3$ 
around any point in space becomes more obvious in cylindrical coordinates 
$(t,r,\phi)$, in which the metric takes the following form, 
\eqn\eegdcyl{ds^2_3=-dt^2+dr^2-\frac{1}{\Omega^2}\left(\sinh^4(\Omega r)
-\sinh^2(\Omega r)\right)d\phi^2-\frac{2\sqrt{2}}{\Omega}\sinh^2(\Omega r)
\,dt\,d\phi.}
Indeed, $\p/\p\phi$ is a Killing vector, of norm squared
\eqn\eegdnorm{\left|\frac{\p}{\p\phi}\right|^2=\frac{1}{\Omega^2}
\left(1-\sinh^2(\Omega r)\right)\sinh^2(\Omega r).}
The orbits of this Killing vector are closed, and become closed timelike 
curves for $r>r_0$,
\eqn\eerzero{r_0=\frac{1}{\Omega}\arcsinh(1)\equiv\frac{1}{\Omega}
\ln(1+\sqrt{2}).}
We will call the surface of $r=r_0$ the {\it velocity-of-light surface}; the 
null geodesics emitted from the origin in this coordinate system reach 
the velocity-of-light surface in a finite affine parameter, and then spiral 
back to the origin where they refocus, again in finite affine parameter.    

\fig{The geometry of the three-dimensional part $\CG_3$ of \godel's 
universe, with the flat fourth dimension $z$ suppressed.  
Null geodesics emitted from the origin $P$ follow a spiral trajectory, reach 
the velocity-of-light surface at the critical distance $r_0$, and spiral back 
to the origin in finite affine parameter.  The curve $C$ of constant $r>r_0$ 
tangent to $\p/\p\phi$ is an example of a closed timelike curve.  A more 
detailed version of this picture appears in Hawking and Ellis 
\hawkell.}{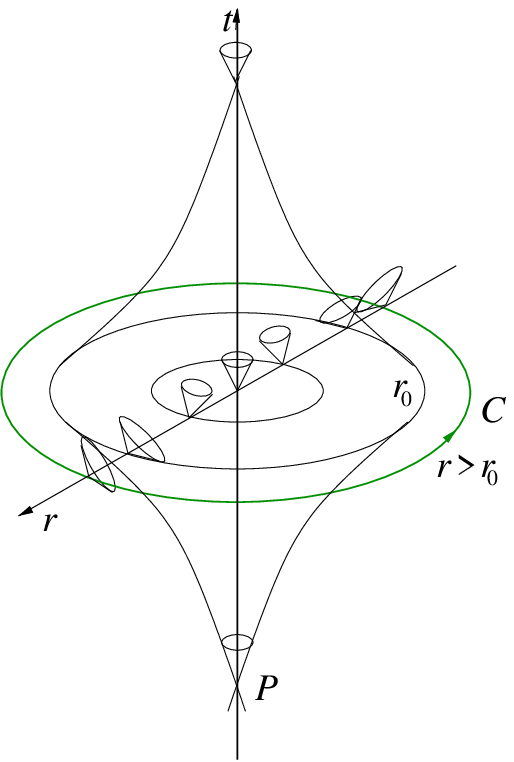}{2.8truein}

The homogeneity of the solution implies that there are closed timelike curves 
through every point in spacetime.  Note that in a well-defined sense all the 
closed timelike curves are topologically ``large'': In order to complete 
a closed timelike trajectory starting at any point $P$, one has to travel 
outside of the velocity-of-light surface (as defined by an observer at $P$) 
before being able to return to $P$ along a causal curve.  This fact will play 
an important role in our argument for 
the holographic resolution of the problem of closed timelike curves below.  
Notice also that none of the closed timelike curves is a geodesic, and that 
the closed timelike curves cannot be trivially eliminated by a lift to the 
universal cover: The manifold is already topologically trivial.  

\godel's universe represents a solution with a good timelike Killing 
vector (indeed, $\p/\p t$ is Killing and everywhere timelike), which however 
cannot be used to define a universal time function: The slices of the 
foliation by surfaces of constant $t$ are not everywhere spacelike.  
The classical Cauchy problem is always globally ill-defined for this 
geometry.  

\subsec{Preferred holographic screens in \godel's universe} 

We now apply Bousso's phenomenological framework for holography 
\refs{\raphael,\lds,\holorev} to \godel's universe.  We indentify its 
preferred holographic screens, associated with particular observers as 
follows:   

Consider a geodesic observer comoving with the distribution of dust in 
\godel's universe (and placed at the origin $r=0$ of our coordinate 
system without loss of generality).  Imagine that the observer sends out 
lightrays in all directions from the origin at some fixed time, say $t=0$.  
These lightrays will at first expand -- i.e., the surfaces that they reach in 
some fixed affine parameter $\lambda$ will grow in area, at least for small 
enough values of $\lambda$.  The preferred holographic screen will be reached 
when we reach the surface of maximal area (or maximal geodesic expansion).   

Alternatively, one can follow {\it incoming} lightrays into their past, until 
reaching the surface where the geodesics no longer expand.  This is again 
the location of the preferred screen $\CB$. The preferred screen $\CB$ can 
then be used to impose a covariant bound on the entropy inside the region 
of space surrounded  by $\CB$ \raphael, which should not exceed one-fourth 
of the area of $\CB$ in Planck units.  

We will first analyze the three-dimensional part $\CG_3$ of \godel's 
solution, which contains much of the nontrivial geometry.  
Even though all the geodesics of \godel's universe are known \chandra , one 
can in fact use the symmetries of $\CG_3$ to determine the location of the 
screen without any explicit knowledge of the geodesic curves.  Since $\CG_3$ 
is rotationally invariant in $\phi$, all the null geodesics emitted from the 
origin will reach the same radial distance $r(\lambda)$ within the same 
affine parameter (assuming that we use a rotationally invariant normalization 
of $\lambda$ for geodesics emitted in different directions from the origin), 
and also for the same global time coordinate $t$.  Thus, to determine the 
surface of maximal geodesic expansion, we can just evaluate the area of the 
surfaces of constant $r$ and $t$ (in our case of course one-dimensional),
\eqn\eeaaone{A=\frac{2\pi}{\Omega}\sinh(\Omega r)\sqrt{1-\sinh^2(\Omega r)},}
and maximize it as a function of $r$.  This very simple calculation yields 
a preferred screen $\CH$ that is isomorphic to a cylinder of constant 
$r=r_\CH$ and any $t$, with
\eqn\eerholo{r_\CH=\frac{1}{\Omega}\arcsinh\left(\frac{1}{\sqrt{2}}\right).}
\fig{The geometry of our preferred holographic screen in \godel's universe, as 
defined by the inertial observer following the comoving geodesic at the 
origin of spatial coordinates.  The translationally invariant dimension $z$ 
is again suppressed. Two closed timelike curves are indicated: One, $C$, at 
constant value of $t=0$ and $r>r_0$ is outside of the preferred screen, while 
another, $C'$, passes through the origin at $t=0$ and intersects the screen 
in two, causally connected points.}{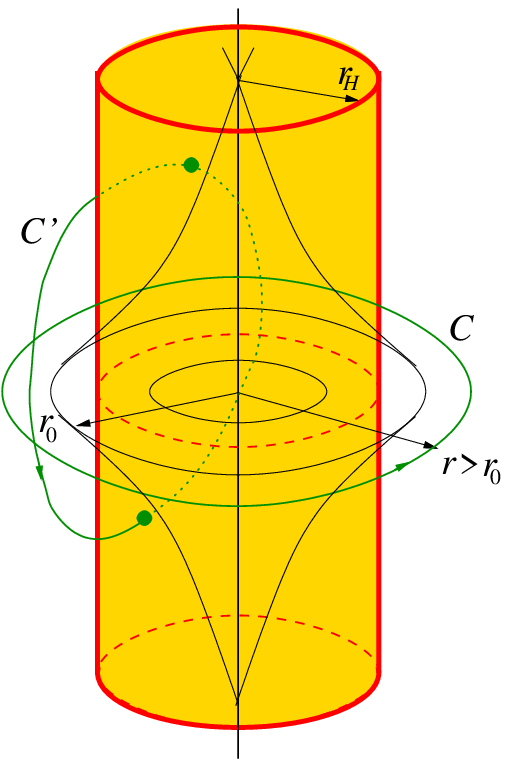}{2.5truein}
Of course, this screen is observer-dependent, in this case associated with 
the comoving inertial observer located at the origin for all values of $t$.  
Other comoving inertial observers would see different but isomorphic screens, 
in a pattern similar to the structure of cosmological horizons associated with 
inertial observers in de~Sitter space.  

One can take advantage of the rotational symmetry of the solution, and 
visualize the location of the preferred screen using a spacetime 
diagram of the type introduced by Bousso \raphael\ (see Figure~3).  This 
diagram suppresses the dimension of rotational symmetry $\phi$, and its points 
represent (in our case one-dimensional) orbits of the rotation group, i.e., 
surfaces of constant $r$ and $t$.  For each such surface, one can define 
the total of four lightsheets: Two oriented forward in time, and two oriented 
backward.  In generic points of the diagram, two of these lightsheets will be 
non-expanding.  At each point of the Bousso diagram one can draw a wedge 
pointing in the direction of non-expanding lightsheets.  These wedges then 
point in the direction of the preferred holographic screen.  
\fig{The Bousso diagram for the $\CG_3$ part of the \godel\ universe metric, 
with the angular coordinate $\phi$ suppressed, and the structure of 
non-expanding lightsheets indicated by the bold wedges.  The preferred 
holographic screen is at the finite value $r_\CH$ of the radial coordinate 
$r$, strictly smaller than the location of the velocity-of-light surface at 
$r_0$. A null geodesic sent from $P$ would reach the velocity-of-light 
surface at $P''$ in a finite affine parameter, and refocus again at the 
spatial origin in $P'$.}{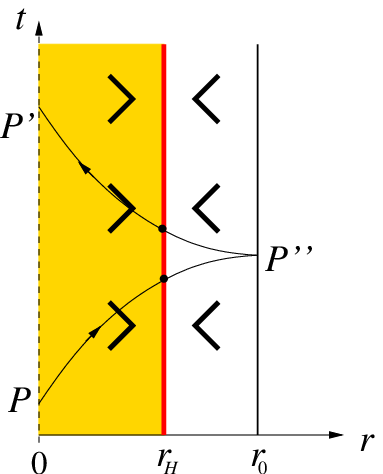}{2.5truein}

One can directly verify that our preferred holographic screen satisfies the 
defining property
\eqn\eezero{\theta=0,}
where $\theta$ is the expansion parameter defined for a spacelike 
codimension-two surface $B$ (in any spacetime with coordinates $x^\mu$) as 
\eqn\eethetadef{\theta=h^{\mu\nu}D_\mu \zeta_\nu,}
with $\zeta_\mu$ the light-like covector orthogonal to $B$ (smoothly but 
arbitrarily extended to some neighborhood of $B$), $D_\mu$ is the covariant 
derivative, and $h_{\mu\nu}$ is the induced metric on $B$.  The most 
convenient way of identifying the surface of $\theta=0$ in \godel's universe 
is to use as $\zeta$ the vector tangent to the congruence of null geodesics 
emitted by the observer at the origin.  An explicit calculation confirms in 
this case that $\theta$ is proportional to $\p_rg_{\phi\phi}$, and therefore 
vanishes at the surface of $r=r_\CH$.  

The metric induced on the preferred holographic screen $\CH$ is of signature 
$(-+)$, everywhere nonsingular:
\eqn\eemetscr{ds^2_\CH=-dt^2+\frac{1}{4\Omega^2}d\phi^2
+\frac{\sqrt{2}}{\Omega}d\phi\,dt}
with $0\leq\phi\leq 2\pi$.  The preferred holographic screen carves out a 
cylindrical compact region of spacetime (which we 
will call the {\it holographic region\/}) in the $\CG_3$ part of \godel's 
universe, centered on the comoving inertial observer at the origin.  This 
region contains no closed timelike curves, as can be easily demonstrated 
by noticing that the causal structure of the holographic region is identical 
to that of a cylindrical portion of (the universal cover of) $AdS_3$.  
The closed timelike curves of the full $\CG_3$ geometry fall into two 
categories: Either they stay completely outside of the holographic region, 
or they enter it and leave it again after traveling a causal trajectory 
within the holographic region.  

\bigskip
\noindent
{\it Preferred screens in $\CG_3\times\R$}
\smallskip

The full \godel\ universe is of the direct product form $\CG_3\times\R$.  
The presence of the extra, translationally-invariant dimension parametrized by 
$z$ actually implies a richer structure of preferred screens than the one 
we just found in the $\CG_3$ factor.  This is in fact a preview of what 
we will find in the next section in the case of supersymmetric \godel\ 
solutions in M-theory and string theory: Those solutions typically also 
contain extra flat dimensions.  

First of all, there is one preferred screen that can be easily identified: 
The three-dimensional surface $\CH\times\R$, where $\CH$ is the preferred 
screen associated with the observer at the spatial origin in $\CG_3$, and 
$\R$ is the extra coordinate $z$, clearly satisfies the zero-expansion 
condition \eezero .  Thus, by definition, this surface $\CH\times\R$ is a 
preferred screen.  This screen is observer-dependent, and the observer 
associated with it can be thought of either as a string wrapped around $z$ or 
as a more traditional observer ``delocalized'' along $z$, each localized at 
the origin of coordinates in $\CG_3$.  Unless we compactify $z$ on $\S^1$, 
the overall area of this translationally-invariant screen is of course 
infinite, but the screen still has a finite ``area density'' per unit 
distance along $z$.  

Alternatively, one can ask what is the preferred screen associated with an 
localized inertial observer in $\CG_3\times\R$.  If one follows null geodesics 
emitted from (or converging onto) a point in $\CG_3\times\R$ where the 
the observer is located, one finds that the surface of maximal geodesic 
expansion is at a finite distance from the observer in all space directions 
including $z$.  This compact, translationally-noninvariant screen is 
completely contained within the velocity-of-light surface as defined by the 
observer.  

For either of these two classes of screens in $\CG_3\times\R$, all closed 
timelike curves are again either hidden outside of the screen or broken by 
it into causal observable pieces.  

\bigskip
\noindent
{\it Covariant entropy bounds and screen complementarity}
\smallskip

The existence of preferred screens, and the structure of the Bousso diagram 
for \godel's universe imply a holographic entropy bound on the amount 
of entropy through any spatial slice of the compact holographic region 
associated with each screen.  This entropy is limited by one fourth of the 
area of the screen measured in Planck units.  Our screen is neither at 
conformal infinity, nor located at a horizon.  The closest analog would be 
the preferred holographic screen located at the equator of the Einstein 
static universe.  Just as in that case, the holographic screen of \godel's 
universe can be used to bound the entropy in either direction normal to the 
screen.  In particular, the lightrays that start at the screen and travel 
in the direction of larger values of $r$ refocus at the velocity-of-light 
surface, and then travel back again to the screen.  This is rather reminiscent 
of the behavior of lightrays in Einstein's static universe: lightrays emitted 
from one pole of the spatial sphere reach the screen at the equator and travel 
to the other hemisphere, refocus at the opposite pole, and travel back to 
the screen and then to the point they were originally emitted from.  

The strong version of the holographic principle suggests that the compact 
holographic screen implies a finite bound on the number of degrees of 
freedom effectively accessible to the inertial observer.  The good causal 
structure of the holographic region associated with that observer may 
suggest that the quantum mechanics of this finite number of degrees of 
freedom could be well-defined, and screened from the acausal behavior outside 
of the velocity-of-light surface by a screen complementarity principle.  

Of course, one may find the very definition of entropy in spacetimes with 
closed timelike curves somewhat problematic.  However, in the case of 
\godel's universe all that matters for our argument is the region strictly 
below the velocity-of-light surface.  One can in principle 
imagine cutting \godel's solution off at some finite $r$ larger than $r_\CH$ 
but smaller than $r_0$, and replacing the outside with some causal geometry.  
The covariant entropy bound can then be safely applied to 
the holographic region, without any possible conceptual difficulties with the 
definition of entropy in the presence of closed timelike curves. 

The intricate structure of compact preferred screens associated with the 
observers in \godel's universe suggests that holography may be the correct, 
causal way of thinking about this geometry without modifying it.  
However, one is forced to replace the naive ``metaobserver'' perspective of 
the geometry by a system of local observers, each of which sees a causal 
region screened from the rest of the naive classical geometry by the preferred 
holographic screen.  Each individual observer would only have access to 
a finite amount of degrees of freedom associated with the corresponding 
holographic region.  Within this finite number of degrees of freedom, 
causality and quantum mechanics would be protected.  

In this paper we will not discuss non-inertial observers attempting to 
travel along closed timelike curves.  In the spirit of Hawking's original 
chronology protection conjecture \chrono , one may expect a large 
backreaction from the geometry that can protect the solution from such 
observers.  

\subsec{\godel's universe as deformed AdS${}_3$ and holography}

It is useful to embed our discussion of \godel's universe into a broader 
framework.  Consider all spacetime-homogeneous metrics of the \godel\ type.   
It has been shown \hommet\ that this family of metrics is parametrized by two 
parameters, $\Omega$ and $m$, with the metric given by
\eqn\eehommet{ds^2=-\left(dt+\frac{4\sqrt{2}\Omega}{m^2}\sinh^2
\left(\frac{mr}{2}\right)d\phi\right)^2+\frac{1}{m^2}\sinh^2(mr)
\,d\phi^2+dr^2+dz^2,}
with $\Omega\in\R$ and $m^2\in\R$.  For $m^2=4\Omega^2$, we recover \godel's 
metric \eegdthree .  On the other hand, for $m^2=8\Omega^2$ we get 
the direct-product metric on $AdS_3\times\R$ \squash .  Notice also that the 
metric simplifies in the limit of $m\rightarrow 0$ keeping $\Omega$ fixed; 
this metric has been analyzed by Som and Raychaudhuri \somr , and is in fact 
a closer analog of the string theory \godel\ universe than \godel's solution 
itself.  

Since all the solutions in \eehommet\ are rotationally invariant, we can 
easily identify the preferred screens for this entire family of metrics 
using the same symmetry argument as in \godel's universe itself.  
The holographic screens $\CH$ of the non-trivial three-dimensional part of 
\eehommet\ are now located at 
\eqn\eehomsc{r_\CH=\frac{2}{m}\arcsinh\left(\left(
\frac{16\Omega^2}{m^2}-2\right)^{-1/2}\right).}
Thus, for $m^2<8\Omega^2$, the screen is at a finite value of $r_\CH$, and 
as we approach the $AdS_3\times\R$ limit it recedes to infinity and becomes 
the canonical holographic screen of $AdS_3$.   This connection with $AdS_3$ 
leads to a particularly intriguing way of thinking about holography of this 
family of solutions in terms of breaking conformal invariance on the 
holographic screen of $AdS_3$ once we move away from the $AdS_3$ limit.  

Clearly, our observation that preferred holographic screens can either screen 
closed timelike curves or break them up into causal pieces is not restricted 
to homogeneous spacetimes.   An example of the same phenomenon in an 
inhomogeneous solution can be easily found: Consider the classic cylindrically 
symmetric inhomogeneous solution with closed timelike curves found in 1937 
by van~Stockum, \stockum , which in the cylindrical coordinates takes the 
form  
\eqn\eestockum{ds^2=-dt^2-2\Omega r^2d\phi\,dt+r^2(1-\Omega^2r^2)d\phi^2+
e^{-\Omega^2r^2}(dz^2+dr^2).}
It is straightforward to show that the preferred holographic screen as 
defined by the inertial observer located at the origin is again compact 
and shields the closed timelike curves from the observer, just as in the 
case of the homogeneous \godel\ universe.  

\newsec{Holography in the Supersymmetric \godel\ Universe}

The \godel\ solution of M-theory found in \jerome\ has a direct product 
form $\CG_5\times\R^6$, where the non-trivial five-dimensional part 
$\CG_5$ represents a maximally supersymmetric solution of minimal supergravity 
in five dimensions.  The underlying spacetime of $\CG_5$ is topologically 
trivial, isomorphic to $\R^5$.  Again, just as in the case of \godel's 
four-dimensional solution, much of the nontrivial structure of the solution 
is carried in this five-dimensional factor $\CG_5$, which plays a role 
analogous to $\CG_3$ of the previous section.  We will therefore study 
holography of this five-dimensional solution first.  

\subsec{Holography in the \godel\ universe of $N=1$ $d=5$ supergravity}

The five-dimensional \godel\ geometry $\CG_5$ is a maximally supersymmetric,  
topologically trivial, homogeneous solution of minimal five-dimensional 
supergravity \jerome .  We introduce generic coordinates $X^\mu$, 
$\mu=0,\ldots 4$ on $\R^5$, but we will soon specialize to several specific 
coordinate systems.  The minimal $d=5$ supergravity contains an Abelian gauge 
field $A_\mu$ whose field strength $F_{\mu\nu}$ we normalize such that the 
Lagrangian has the following form, 
\eqn\eefsugra{\CL_5=\frac{1}{2\kappa^2_5}\int d^5X\left(R-\frac{1}{4}F_{\mu\nu}
F^{\mu\nu}+\ldots\right),}
where the ``$\ldots$'' stand for a Chern-Simons self-interaction of the 
gauge field and for fermionic terms.  

The \godel\ solution takes the form of a fibration over the flat Euclidean 
$\R^4$ with fibers isomorphic to $\R$ and with a simple twist, which in a 
Cartesian coordinate system $t,x_i$, $i=1,\ldots 4$, can be written as 
\eqn\eefiveg{\eqalign{ds^2&=-(dt+\beta\omega)^2+\sum_{i=1}^4dx_i^2,\cr
F&=2\sqrt{3}\beta J,\cr}}
with the twist one-form $\omega$ given by 
\eqn\eefiveomega{\omega=x_1dx_2-x_2dx_1+x_3dx_4-x_4dx_3\equiv J_{ij}x_idx_j,}
and $J_{12}=-J_{21}=J_{34}=-J_{43}=1$ a preferred K\"ahler form on $\R^4$.   
In \eefiveg , $\beta$ is a constant rotation parameter, of mass dimension 
one.  Without any substantial loss of generality, we will assume $\beta$ to be 
positive.  

As remarked in \jerome , this solution is homogeneous, and in fact has a 
nine-dimensional group of bosonic isometries.  The Killing vectors are given 
by 
\eqn\eekillvectcomp{\eqalign{P_0&=\p_t,\cr
P_i&=\p_i-\beta J_{ij}x_j\p_t,\cr
L&=x_1\p_2-x_2\p_1+x_3\p_4-x_4\p_3,\cr
R_1&=x_1\p_2-x_2\p_1-x_3\p_4+x_4\p_3,\cr
R_2&=x_1\p_3-x_3\p_1+x_2\p_4-x_4\p_2,\cr
R_3&=x_1\p_4-x_4\p_1+x_3\p_2-x_2\p_3,\cr
}}
where $\p_i=\p/\p x_i$.  The commutation relations of this bosonic symmetry 
algebra are 
\eqn\eecomm{\eqalign{[R_\alpha,R_\beta]&=2\epsilon_{\alpha\beta\gamma}R_\gamma,
\qquad\qquad [L,R_\alpha]=0,\cr
[P_i,P_j]&=2\beta J_{ij}P_0.\cr}}
Here $\alpha,\beta,\ldots=1,2,3$ go over a basis of anti-selfdual two-tensors 
in $\R^4$.  $R_\alpha$ and $L$ act on the momenta $P_i$ as rotations.  
Thus, we find that the symmetry algebra of the \godel\ universe 
$\CG_5$ is given by the semidirect product $H(2)\semi(SU(2)\times U(1))$, 
where $H(2)$ is the Heisenberg algebra with five generators.%
\foot{As we will see in Section~4, the remarkable similarity between this 
symmetry algebra and a pp-wave symmetry algebra is not a coincidence:  When 
lifted to string theory, the \godel\ solution is actually T-dual to a 
supersymmetric pp-wave!  Notice, however, that in the symmetry algebra of 
$\CG_5$, the central extension generator $P_0$ of the Heisenberg algebra is 
represented by a timelike Killing vector, while in the pp-wave it would be 
null.  One can actually show by a direct calculation that the five-dimensional 
\godel\ universe (or the string theory lifts thereof to be studied below) 
does not admit any covariantly constant null vectors, which proves that 
it is not ``secretly'' a pp-wave in unusual coordinates.}

While the translation symmetries $P_i$ of the solution are almost manifest in 
the cartesian coordinates $t,x_i$, the rotation symmetries are rather 
obscure.  It is therefore convenient to introduce a new coordinate system.  
First, we introduce a pair of polar coordinates, one in each of the two main 
planes of rotation, 
\eqn\eefivebipol{\eqalign{x_1=r_1\cos\phi_1,\cr
x_2=r_1\sin\phi_1,\cr}\qquad
\eqalign{x_3=r_2\cos\phi_2,\cr
x_4=r_2\sin\phi_2.\cr}}
In these ``bipolar'' coordinates, the metric becomes
\eqn\eemetbipol{\eqalign{ds^2&=-dt^2-2\beta(r_1^2d\phi_1+r_2^2d\phi_2)dt
+dr_1^2+dr_2^2-2\beta^2r_1^2r_2^2\,d\phi_1d\phi_2\cr
&\qquad\qquad{}+r_1^2(1-\beta^2r_1^2)d\phi_1^2+
r_2^2(1-\beta^2r_2^2)d\phi_2^2.\cr}}
The non-Abelian part of the rotation symmetries becomes manifest in 
spherical coordinates $(r,\phi_1,\phi_2,\vartheta)$, with $\vartheta
\in[0,\pi/2)$, 
\eqn\eefivepol{\eqalign{x_1+ix_2&=r\,e^{i\phi_1}\cos\vartheta,\cr
x_3+ix_4&=r\,e^{i\phi_2}\sin\vartheta,\cr}}
which bring the metric to the following form,
\eqn\eemetsph{ds^2=-\left(dt+\frac{\beta r^2}{2}\sigma_3\right)^2+dr^2+
r^2d\Omega_3^2.}
Here $d\Omega_3^2$ is the standard unit-volume metric on $\S^3$, and 
$\sigma_3$ is one of the right-invariant one-forms on $SU(2)$, 
\eqn\eesigthree{\sigma_3=2(\cos^2\vartheta\,d\phi_1+\sin^2\vartheta\,d\phi_2).}
It is clear from this expression for the metric that even though the solution 
does not exhibit the full $SO(4)\sim SU(2)\times SU(2)$ rotation symmetry in 
$\R^4$, the non-zero rotation parameter $\beta$ keeps the right $SU(2)$ 
(together with a $U(1)$ subgroup of the left $SU(2)$) unbroken.  

It was also noted in \jerome\ that the \godel\ universe $\CG_5$ preserves 
all eight supersymmetries of minimal $d=5$ supergravity.  Thus, the bosonic 
symmetry algebra \eecomm\ will extend to a superalgebra with eight 
supercharges $Q$.  It is natural to split $Q$ into two four-component 
spinors, $Q^\pm$.  In this notation, the (anti)commutation relations of the 
full symmetry superalgebra can be written as follows, 
\eqn\eeanticomm{\eqalign{[P_0,Q^\pm]&=0,\cr
[P_i,Q^+]&=0,\cr
\{\bar Q^+,Q^+\}&=\Gamma^0P_0,\cr
\{\bar Q^-,Q^-\}&=\Gamma^0(P_0+2\beta L),\cr
}\qquad
\eqalign{[R,Q^\pm]&=\Gamma_R Q^\pm,\cr
[P_i,Q^-]&=\beta J_{ij}\Gamma^{j}Q^+,\cr
\{\bar Q^-,Q^+\}&=\Gamma^iP_i,\cr
&\cr}}
together with \eecomm.  In \eeanticomm, $R$ denotes any of the rotation 
generators $R_\alpha$ or $L$, and $\Gamma_R$ is a shorthand for the generator 
of conventional rotations associated with $R\in SO(4)$, in the corresponding 
spinor representation of $SO(4)$.  

Once we examine the structure of preferred holographic screens in the next 
subsection, it will be interesting to see how these screens are compatible 
with the structure of the supersymmetry algebra \eecomm , \eeanticomm .  

\subsec{Preferred holographic screens}

Consider an inertial, comoving observer located at an arbitrary point in 
space, which we place without any loss of generality at the origin of 
cartesian coordinates $x_i=0$.  Since we are focusing on the perspective 
of an observer at the origin, it will be convenient to use either the 
``bipolar'' or the spherical coordinates.  

The symmetry arguments that allowed us to identify the preferred screen in 
\godel's universe $\CG_3$ without actually calculating the geodesics 
can in fact be extended to the supersymmetric solution $\CG_5$ as well.  
Despite the fact that the full $SO(4)$ rotation symmetry of $\R^4$ 
is broken to an $SU(2)\times U(1)$ subgroup, the unbroken group still acts 
transitively on the three-spheres of constant $r$.  Indeed, one can think 
of the $\S^3$ at constant $r$ as a copy of $SU(2)$, on which the full 
$SO(4)$ rotations would act by the left action of one $SU(2)$ and 
the right action of the other $SU(2)$.  In the \godel\ solution, the metric 
on the $\S^3$ of constant radius is that of a squashed three-sphere, which 
still leaves the (transitive) right action by $SU(2)$ unbroken.  
This unbroken $SU(2)$ is sufficient to reduce our analysis of the location 
of preferred screens to the maximization of the area of the surfaces $\S^3$ 
of constant $r$ as a function of $r$ (at constant $t$), precisely as in the 
simpler case of $\CG_3$ studied in the previous section.  Without knowing the 
precise structure of the null geodesics emitted at some time $t<0$ in all 
directions from the origin, the symmetries imply that these geodesics will 
reach the $\S^3$ of some fixed radius $r$ at $t=0$.  

Thus, in order to find the preferred holographic screens associated with the 
inertial comoving observer at the origin, we only need to maximize the volume 
of the $\S^3$ at fixed $r$, as a function of $r$.  The induced metric on the 
$\S^3$ of radius $r$ at constant $t$ is given by
\eqn\eeindmet{\eqalign{ds^2_{\rm ind}&=r^2d\vartheta^2+r^2\cos^2\vartheta
(1-\beta^2r^2\cos^2\vartheta)d\phi_1^2\cr
&\qquad{}+r^2\sin^2\vartheta(1-\beta^2r^2\sin^2\vartheta)d\phi_2^2-
2\beta^2r^4\cos^2\vartheta\sin^2\vartheta\,d\phi_1d\phi_2,\cr}}
implying that the induced area of this surface is given by
\eqn\eearear{\CA(r)=\int_{\S^3}\sqrt{h_{\rm ind}}=2\pi^2
r^3\sqrt{1-\beta^2r^2},}
where $h_{\rm ind}$ is the determinant of the induced metric \eeindmet .  
We conclude that the preferred holographic screen is located at radial 
distance $r$ (call it $r_\CH$) where the area \eearear\ is maximized, 
\eqn\eesgdlrh{r_\CH=\frac{\sqrt{3}}{2\beta}.}
The screen carries a Lorentz-signature induced metric, 
\eqn\eescrmet{\eqalign{ds^2_\CH&=-dt^2-\frac{3}{2\beta}(\cos^2\vartheta\,
d\phi_1
+\sin^2\vartheta\,d\phi_2)\,dt+\frac{3}{4\beta^2}\left[\vphantom{\frac{1}{2}}
d\vartheta^2+\cos^2\vartheta\,d\phi_1^2+
\sin^2\vartheta\,d\phi_2^2\right.\cr
&\qquad\qquad\qquad\qquad\left.{}-\frac{3}{4}(\cos^2\vartheta\,
d\phi_1+\sin^2\vartheta\,d\phi_2)^2\right],\cr}}
with each spacelike slice of constant $t$ 
isomorphic to the squashed three-sphere of radius $r_\CH$ and squashing 
parameter $3/4$.  The screen metric \eescrmet\ seems to exhibit dragging 
of frames, but this is an artifact of a coordinate choice.  Upon intoducing 
new angular coordinates by $\bar\phi_1=\phi_1-4\beta t$, 
$\bar\phi_2=\phi_2-4\beta t$, \eescrmet\ becomes
\eqn\eescrnew{ds^2_\CH=-4dt^2+\frac{3}{4\beta^2}\left[
d\vartheta^2+\cos^2\vartheta\,d\bar\phi_1^2+
\sin^2\vartheta\,d\bar\phi_2^2-\frac{3}{4}(\cos^2\vartheta\,
d\bar\phi_1+\sin^2\vartheta\,d\bar\phi_2)^2\right].}
This phenomenon is analogous to the behavior of horizons in rotating 
black holes in five dimensions \townsend . 

The screen and its location in the \godel\ universe 
can be visualized exactly as in Fig.~2, with $\phi$ now collectively denoting 
the coordinates on the squashed three-sphere.  Again, the preferred screen 
cuts out a compact region of space with the observer inside, which we will 
refer to as the holographic region.  

The compact preferred holographic screen also implies a finite bound 
on the entropy that flows through a space-like section of the holographic 
region.  This entropy has to be smaller than one fourth of the area of the 
screen in Planck units,
\eqn\eefivebound{S\leq\frac{2\pi^3 r_\CH^3}{\kappa_5^2}.}
(Notice that our $\kappa_5$ is related to the 5d Newton constant by 
$8\pi G_N=\kappa_5^2$.) 

It is interesting to analyze the symmetries preserved by the screen.  
While all the rotation symmetries $SU(2)\times U(1)$ as well as the time 
translation symmetry are left unbroken, all the space translations are 
broken by the screen.  Similarly, the structure of the supersymmetry 
algebra reveals that one half of the supercharges (namely $Q^-$) will be 
broken by the screen, while the remaining half of supersymmetry represented 
by $Q^+$ (and associated with Killing spinors which are simply constant) 
is compatible with the presence of the screen.  Thus, the 
screen can preserve as much as 1/2 of the full supersymmetry of the 
\godel\ solution, leaving an unbroken symmetry which coincides with the 
symmetry left unbroken by the choice of the inertial comoving observer.  
Once we lift the solution to M-theory, we can also think of the preferred 
comoving observer as a massless particle moving with the speed of light along 
the extra dimension and preserving 1/2 of supersymmetry.  Thus, the 
symmetries of the observer seem compatible with the symmetries that can 
be left unbroken by her preferred holographic screen. 

In order to verify that this simplified argument for identifying the preferred 
screens, which relies on the large symmetry of the solution, coincides with 
the conventional local definition of the screen \raphael\ as the surface of 
vanishing expansion parameter $\theta=0$ of the null geodesics emitted from 
(or, by the time reflection symmetry, sent towards) the origin in space, we 
must first analyze the structure of geodesic motion in the \godel\ spacetime.  
This analysis will also refine our understanding of the \godel\ universe 
geometry.  

\subsec{Geodesics in the \godel\ universe $\CG_5$}

In this subsection we will find all the geodesics in the \godel\ universe.  

First, one can use the symmetries of the solution to simplify the 
analysis.  By homogeneity, it will be sufficient to consider geodesics through 
the origin $P$ of our coordinate system, $P\equiv\{t=x_m=0\}$.  In any case, 
for the identification of the preferred screens we are primarily interested 
in null geodesics emitted from the origin.%
\foot{Moreover, since the $SU(2)$ part of the symmetry group acts 
transitively on the celestial sphere at $P$, one could rotate the initial 
momentum vector along the geodesic to lay entirely in the $x_3=x_4=0$ plane.  
By angular momentum conservation, corresponding to the two Killing vectors 
$\p/\p\phi_1$ and $\p/\p\phi_2$, the geodesic would then stay in the 
$x_3=x_4=0$ plane throughout its history.} 

We will write the tangent vector to the geodesic as 
\eqn\eetangeo{\xi=\dot t\frac{\p}{\p t}+\dot r_1\frac{\p}{\p r_1}+
\dot\phi_1\frac{\p}{\p\phi_1}+\dot r_2\frac{\p}{\p r_2}+
\dot\phi_2\frac{\p}{\p\phi_2},}
where $\dot{}\equiv d/d\lambda$ denotes the derivative with respect to an 
affine parameter $\lambda$ along the geodesic.  

The large amount of symmetry of the \godel\ universe allows us to explicitly 
solve for all the geodesics without any restrictions.  First of all, the 
following integrals of motion will be useful, 
\eqn\eeintmot{\eqalign{(\xi,\xi)&=-M^2,\qquad (\xi,\p_t)=-E,\cr
(\xi,\p_{\phi_1})&=L_1,\qquad (\xi,\p_{\phi_2})=L_2.\cr}}
Here $L_1,L_2$ are the angular momenta in the two preferred planes of 
rotation.  The $\pm$ sign of $M^2$ corresponds to timelike and spacelike 
geodesics, with $E$ the energy of the particle in the timelike case.  In the 
null case $M^2=0$ we will find it convenient to rescale the affine parameter 
$\lambda$ along the geodesic so as to set $E=1$.  

The integrals of motion \eeintmot\ imply
\eqn\eeinteom{\eqalign{\dot\phi_1&=\beta E+\frac{L_1}{r_1^2},\qquad\qquad
\dot\phi_2=\beta E+\frac{L_2}{r_2^2},\cr
\dot t&=(1-\beta^2r_1^2-\beta^2r_2^2)E-\beta(L_1+L_2),\cr}}
as well as 
\eqn\eeeomsqu{(\dot r_1)^2+(\dot r_2)^2-(1-\beta^2r_1^2-\beta^2r_2^2)E^2
+2\beta E(L_1+L_2)+\frac{L_1^2}{r_1^2}+\frac{L_2^2}{r_2^2}=-M^2.}

In order to identify the holographic screen we need the null geodesics 
going through the origin.  Note that for non-zero values of the angular 
momenta $L_1$ or $L_2$, the effective potential for $r_1$ and $r_2$ precludes 
the geodesics from reaching the origin $r_1=r_2=0$.  Thus, all the geodesics 
passing through the origin will have $L_1=L_2=0$, and we focus on those now.  
\foot{Of course, all the geodesics with nonzero angular momenta can be easily 
obtained from those with zero angular momenta by the action of the large 
isometry group of the \godel\ metric.}
In order to separate $\dot r_1$ from $\dot r_2$ we need one more integral 
of motion.  Consider 
\eqn\eeextraint{\eqalign{(\xi,R_3)&\equiv(\sin\phi_1\,\sin\phi_2+\cos\phi_1\, 
\cos\phi_2)\left(\frac{r_2}{r_1}L_1+\frac{r_1}{r_2}L_2\right)\cr
&\qquad\qquad{}+(\sin\phi_1\,\cos\phi_2-\cos\phi_1\,\sin\phi_2)(r_2\dot r_1-r_1
\dot r_2).\cr}}
At zero angular momentum, \eeextraint\ has to vanish, implying that 
the angle $\vartheta$ between $r_1$ and $r_2$ is another integral of motion.  
Thus, the equations of motion for the geodesics that pass through the origin 
of space simplify to
\eqn\eeorigeom{(\dot r)^2+\beta^2r^2E^2=E^2-M^2,}
plus \eeinteom\ with $L_i$ set to zero.  These can be easily solved, yielding 
\eqn\eegeods{\eqalign{r_1&=\frac{1}{\beta}\sqrt{1-M^2}\sin(\beta\lambda)
\cos\vartheta,\cr
r_2&=\frac{1}{\beta}\sqrt{1-M^2}\sin(\beta\lambda)\sin\vartheta,\cr
t&=\frac{1}{2}(1+M^2)\lambda+\frac{1}{4\beta}(1-M^2)\sin(2\beta\lambda)+t_0,\cr
\phi_1&=\beta\lambda+\phi_1^{(0)},\cr
\phi_2&=\beta\lambda+\phi_2^{(0)}.\cr}}
with $\vartheta\in[0,\pi/2)$ and $\phi_1^{(0)},\phi_2^{(0)}\in[0,2\pi)$ all 
constants.  We have rescaled the affine parameter $\lambda$ so as to set $E$ 
equal to one.   For null geodesics, $M^2=0$, while for the timelike geodesics 
$M^2\in[0,1]$ as a result of our rescaling of $\lambda$.  Notice that the 
comoving time $t$ at the origin (the coordinate corresponding to the Killing 
vector $\p_t$) is {\it not\/} a good affine parameter along the null 
geodesics passing through the origin.  Instead, either one of the two main 
rotation angles $\phi_1$, $\phi_2$ plays the role of a natural affine 
parameter (as long as $\beta$ is nonzero of course).  

Even though the spherical coordinate system is not smooth at the origin, 
it is easy to verify -- by switching to the original Cartesian 
coordinate system -- that the system of null geodesics \eegeods\ represents 
the complete system of all geodesics passing through the origin.  Indeed, the 
tangent vector to this congruence at $\lambda=0$ is given in the Cartesian 
coordinates by
\eqn\eeinitgeod{\eqalign{\left.\xi\right|_{\lambda=0}=\frac{\p}{\p t}
&+\cos\vartheta\,\cos\phi_1^{(0)}\frac{\p}{\p x_1}
+\cos\vartheta\,\sin\phi_1^{(0)}\frac{\p}{\p x_2}\cr
&\qquad\qquad\qquad{}+\sin\vartheta\,\cos\phi_2^{(0)}\frac{\p}{\p x_3} 
+\sin\vartheta\,\sin\phi_2^{(0)}\frac{\p}{\p x_4},\cr}}
demonstrating that the constants $\vartheta$, $\phi_1^{(0)}$ and 
$\phi_2^{(0)}$ are indeed parametrizing the entire celestial sphere at the 
origin.  

Thus, we see an interesting refocusing behavior of all geodesics in the 
\godel\ universe: 

They start moving from the origin towards larger values of $r$, which at 
first means larger proper-radius spheres, but then at affine parameter 
\eqn\eefocone{\lambda=\frac{\pi}{2\beta}}
they reach the velocity-of-light surface, located at the largest value $r_0$ 
of the radial coordinate $r$ that is accessible by geodesic motion from the 
origin, 
\eqn\eefoctwo{r_0=\frac{1}{\beta}.}
By that time, both $\phi_1$ and $\phi_2$ change exactly by $\pi/2$.  Then it 
takes another 
\eqn\eefocthree{\Delta\lambda=\frac{\pi}{2\beta}}
to complete one period of oscillation and refocus at the origin.  The amount 
of global comoving time coordinate elapsed during the completion of one 
oscillation cycle equals 
\eqn\eefocfour{\Delta t=\frac{\pi}{2\beta}.}
Note that the lightray arrives with its momentum equal to the intial-value 
momentum; thus, the lightray traveled a full circle in the $(x_1,x_2)$ 
plane.  The same holds true for the $(x_3,x_4)$ plane.  

During one refocusing cycle, the proper area of the three-sphere reached by 
the geodesics reaches a maximum twice, precisely when they reach the preferred 
screen -- first on their way out towards the velocity-of-light surface 
(where the proper area of the $\S^3$ goes to zero) and then again on their 
way back to the origin.  In fact, they reach the holographic screen for 
the first time at affine parameter 
\eqn\eefocfive{\lambda=\frac{\pi}{3\beta},}
one third into the oscillation cycle.  

\fig{The behavior of null geodesics emitted from an arbitrary point $P$ 
in the \godel\ universe, with the intial momentum in the $(x_1,x_2)$ 
plane, and with several such geodesics indicated.  Each geodesic travels along 
a circular trajectory, reaches the velocity-of-light surface and returns back 
to $P$, penetrating the preferred screen exactly twice during each rotation 
cycle.}{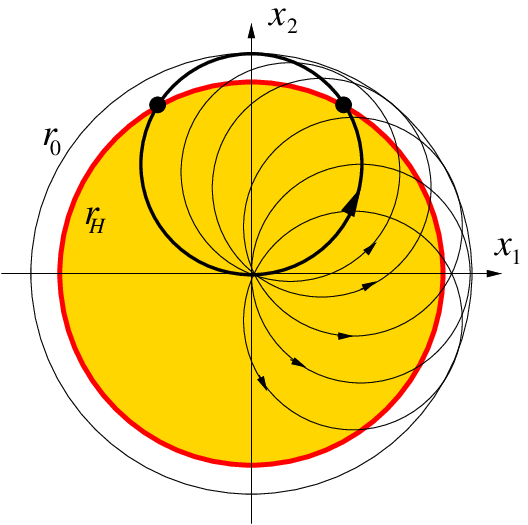}{3truein}

Since any given geodesic moves around a circle in each of the preferred 
planes of rotation, it is instructive to use the translation symmetries of 
the solution, and transform \eegeods\ into the frame associated with the 
observer at the center of this circular motion.  The Killing vectors 
\eekillvectcomp\ can be easily integrated to give finite translations.  For 
example, we find that a finite translation by $a$ along $x_2$ is accompanied 
by an $x_1$-dependent translation in $t$,
\eqn\eefintran{x'_2=x_2+a,\qquad x'_1= x_1,\qquad t'= t+\beta x_1a.}
When one transforms \eegeods\ to the primed coordinates associated with the 
center of the circular motion of a geodesic, the $x_1$-dependent time 
translation \eefintran\ eliminates the $\sin(2\beta\lambda)$ term in the 
expression for $t$ as a function of the affine parameter in \eegeods .  
Thus, $t$ becomes a good affine parameter precisely for the class of 
geodesics that circle around the origin at fixed constant $r$.  

So far, we were mainly concentrating on null geodesics emanating from the 
origin.  The analysis is easily extended to timelike geodesics, which 
turn out to exhibit a similar cyclic behavior.  However, they only 
reach up to a certain critical distance $r_M$ strictly smaller than the 
distance $r_0$ of the velocity-of-light surface, 
\eqn\eecritrmass{r_M=\sqrt{1-M^2}\,r_0.}
In terms of the global comoving time coordinate $t$, the timelike geodesics 
sent from the origin take longer to refocus at the origin than null 
geodesics, the refocusing time being 
\eqn\eerefocmass{\Delta t(M)=\frac{(1+M^2)\pi}{2\beta}.}

\bigskip
\noindent{\it The geodesic expansion $\theta$}
\smallskip

We are now in a position to verify that the holographic screen is indeed 
located at $r_\CH=\sqrt{3}/2\beta$ by a direct analysis of the geodesics 
in the \godel\ metric.  Recall that according to Bousso's prescription 
\raphael , the screen is determined as the surface $\CB$ where the geodesic 
expansion $\theta$ vanishes, leading to the ``equation of motion'' for the 
preferred holographic screen, 
\eqn\eescreenth{\theta=0,}
with $\theta\equiv h^{\mu\nu}D_\mu\xi_\nu$ defined as the contraction of the 
covariant derivative $D_\mu\xi_\nu$ of the null covector $\xi_\mu$ with 
respect to the induced metric $h_{\mu\nu}$ on $\CB$.  

The null geodesics \eegeods\ define a congruence whose associated tangent 
vector is
\eqn\eetangcong{\xi=(1-\beta^2r^2)\frac{\p}{\p t}+\beta\left(
\frac{\p}{\p\phi_1}+\frac{\p}{\p\phi_2}\right)+\sqrt{1-\beta^2r^2}
\frac{\p}{\p r}.}
Its covector dual (which we denote by the same letter $\xi$) has a rather 
simple form,
\eqn\eecovcong{\xi\equiv\xi_\mu dX^\mu=-dt+\sqrt{1-\beta^2r^2}dr.}
We can now evaluate the covariant derivative $D_\mu\xi_\nu$ and contract it 
against the induced metric $h^{\mu\nu}$, to obtain the geodesic expansion 
$\theta$.  After some straightforward algebra,  
\eqn\eethetafive{\theta=\frac{3-4\beta^2r^2}{r\sqrt{1-\beta^2r^2}}.}
Thus, $\theta$ vanishes precisely at $r=r_\CH\equiv\sqrt{3}/2\beta$, in 
accord with our anticipation in \eesgdlrh .  Notice also that $\theta$ 
diverges at the origin and at the velocity-of-light surface, confirming 
that those are indeed caustics of the geodesic motion.  

\subsec{The \godel\ universe of M-theory}

The lift of the five-dimensional \godel\ universe $\CG_5$ to M-theory 
involves adding six flat dimensions $\R^6$, which we parametrize by 
coordinates $y_a$, $a=1,\ldots 6$.  Together, $t,x_i$ and $y_a$ form a 
coordinate system $X^M$ on $\R^{11}$, with $M=0,\ldots 10$.  The action of 
eleven-dimensional supergravity has the form 
\eqn\eeelela{\CL_{11}=\frac{1}{2\kappa^2}\int d^{11}X\left(R-\frac{1}{48}
G_{MNPQ}G^{MNPQ}+\ldots\right),}
where ``$\ldots$'' stand for the Chern-Simons term plus fermionic terms.  
The eleven-dimensional \godel\ solution is then given by
\eqn\eeeleveng{\eqalign{ds^2&=-(dt+\beta\omega)^2+\sum_{i=1}^4dx_i^2+
\sum_{a=1}^6dy_a^2,\cr
G_{ijab}&=2\beta J_{ij}K_{ab},\cr}}
with all the other non-zero components of $G_{MNPQ}$ related to \eeeleveng\ by 
permutations of indices, and the K\"ahler form $K$ on the $\R^6$ factor 
defined by $K_{12}=-K_{21}=K_{34}=-K_{43}=K_{56}=-K_{65}=1$.  

Consider again the congruence of all null geodesics emitted from the origin 
in space, where our comoving observer is located.  The longitudinal 
momenta $K^a$ along $y_a$ are conserved, leading to the following 
congruence of null geodesics:
\eqn\eenullgeoz{\eqalign{r_1&=\frac{1}{\beta}\sqrt{1-K^2}\sin(\beta\lambda)
\cos\vartheta,\cr
r_2&=\frac{1}{\beta}\sqrt{1-K^2}\sin(\beta\lambda)\sin\vartheta,\cr
t&=\frac{1}{2}(1+K^2)\lambda+\frac{1}{4\beta}(1-K^2)\sin(2\beta\lambda)+t_0,\cr
\phi_1&=\beta\lambda+\phi_1^{(0)},\cr
\phi_2&=\beta\lambda+\phi_2^{(0)},\cr
y^a&=K^a\lambda.\cr}}

Just as in the case of four-dimensional \godel's solution $\CG_3\times\R$ 
discussed in the previous section, one can use geodesics in the supersymmetric 
\godel\ solution $\CG_5\times\R^6$ of M-theory to define several different 
classes of preferred screens.  First of all, there is a preferred screen 
which is a direct product of $\R^6$ and the screen that we found at $r=r_\CH$
in $\CG_5$.  This screen is translationally invariant along all the extra 
dimensions $y_a$, and clearly satisfies the $\theta=0$ condition 
trivially.  It is observer-dependent, and should be associated with an 
observer localized at a point in $\CG_5$ but otherwise delocalized along 
$\R^6$, or with the maximal expansion of lightrays sent with zero momentum 
$K^a$ from the origin in $\CG_5$ and arbitrary $y_a$.  

In addition, observers localized in a point $P$ both in $\CG_5$ and in 
$\R^6$ will naturally see a compact screen in all directions.  The precise 
location of this compact screen can be found by considering the full 
congruence \eenullgeoz\ of gedesics emitted from $P$.  One can in principle 
calculate the expansion parameter $\theta$ and find the preferred compact 
screen as the surface of maximal expansion.  
\fig{The two types of preferred screens in the M-theory \godel\ 
$\CG_5\times\R^6$.  
The translationally-invariant screen is located at $r_\CH$ in $\CG_5$ for 
all values of $|y|$, and can be associated with an extended observer 
delocalized or wrapped along $y_a$.  The screen associated with a localized 
observer is compact in all space directions, and extends beyond $r_\CH$, 
closer to the velocity-of-light surface $r_0$.}{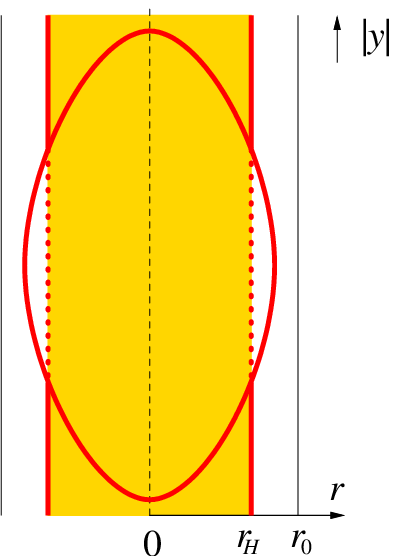}{2.2truein}
Using the affine parameter $\lambda$ and the total momentum $K^2\equiv K_aK^a$ 
along $\R^6$ as coordinates, the shape of the screen is determined from the 
$\theta=0$ condition by a rather complicated implicit function of $\lambda$ 
and $K^2$,  
\eqn\eenice{\eqalign{0&=\frac{1}{2\lambda}\sin^{-1}(\beta\lambda)\left[
(1-K^2)\beta\lambda\,\cos(\beta\lambda)+K^2\,\sin(\beta\lambda)\right]^{-1}\cr
&{}\qquad\qquad\times\left[5K^2+2(1-K^2)\beta^2\lambda^2+(-5K^2+4(1-K^2)
\beta^2\lambda^2)\,\cos(2\beta\lambda)\right.\cr
&\qquad\qquad\qquad\qquad\qquad{}\left.+2(3-K^2)\beta\lambda\,
\sin(2\beta\lambda)\right].\cr}}
This screen is compact in all space dimensions, and exhibits 
$SO(6)\times SU(2) \times U(1)$ rotation invariance, with $SO(6)$ acting 
on $y_a$ and $SU(2)\times U(1)$ on $x_i$.  
  
There are several interesting points about this compact screen.  First of 
all, along $y_a=0$ this screen extends in the $r$ directions beyond the 
location $r_\CH$ of the translationally invariant screen.  This is in fact 
intuitively clear: once we add the flat dimensions $y_a$, the tendency of the 
geodesics to expand in the $y_a$ dimensions competes against the refocusing 
behavior of the geodesics in the $r$ direction of $\CG_5$, effectively slowing 
down the process of reaching the surface of maximal area, which now happens 
for a slightly larger value of $r$.  Notice also that the entire compact 
screen still fits nicely within the velocity-of-light surface 
as defined by our observer.  Therefore, closed timelike curves are again 
shielded from the observer by this screen.  

\newsec{Analogies with Holography in de~Sitter Space}

Holography in de~Sitter space is difficult due to the absence of a solvable 
model or an explicit embedding of de~Sitter into string theory.  As we have 
seen in the previous sections, holography in the \godel\ universes 
exhibits notable analogies with holography in de~Sitter space.  

There are two important classes of preferred holographic screens in de~Sitter 
\raphael : 
First, the future and past infinity are global, observer-independent screens 
of Euclidean signature.  An attempt to formulate holography using these 
screens \edds\ has led to the conjectured dS/CFT correspondence \andyds .  
However, it is difficult to associate these global screens with an observer 
inside de~Sitter: Distinct points at future infinity in de~Sitter are outside 
of each other's causal influence, and any operational definition of measurable 
correlations seems to require a metaobserver.  

The second class of screens is more suitable for the description of physics 
as seen by an observer inside de~Sitter \refs{\tomds,\lds,\petrds}:  
The preferred screen of a given observer is located at his or her 
cosmological horizon.  Since the area of this observer-dependent screen is 
finite, the strong version of the holographic principle implies a finite 
number of degrees of freedom in the quantum mechanics associated with that 
observer.  The finiteness of the number of degrees of freedom accessible to 
any given observer leads to various conceptual puzzles, such as the recently 
discussed question of time recurrences \recurr .  Observers following 
different trajectories have access to different holographic regions, perhaps 
suggesting a quantum mechanical description of de~Sitter space as a web of 
infinitely many Hilbert spaces  (each associated with an observer and 
grasping a finite number of degrees of freedom) with a complicated system 
of maps between them (reflecting the exchange of data between causally 
connected observers, and the horizon complementarity principle).  

Given the conceptual complexity of de~Sitter holography, it would be very 
helpful to have an explicit simple solvable model exhibiting similar 
properties.  We believe that the supersymmetric \godel\ universes may provide 
such a model.  
Indeed, preferred screens appearing in \godel\ holography share many 
properties with the second type of preferred screens in de~Sitter space: 

\item{$\bullet$} Both represent an example of homogeneous geometries with 
screens that are only defined when an observer has been selected.  Observers 
following different worldlines will see different holographic screens.  

\item{$\bullet$} 
The underlying spacetime geometry is homogeneous, but this homogeneity is 
broken by the selection of the observer, and consequently by the location of 
the observer-dependent holographic screen, implying that the screen breaks 
spontaneously some of the symmetries of the naive vacuum. This picture 
of observer-dependent holography stresses the importance of a local, 
environmentally-friendly definition of cosmological observables.  

\item{$\bullet$} The finite proper area of the holographic screen implies a 
finite bound on the entropy that flows through the compact holographic region 
of space associated with the observer.  In addition, the strong version of 
the holographic principle suggests that the observer has only access to a 
finite number of degrees of freedom.  Since the volume of space accessible to 
the observer is effectively finite, the system has effectively been put in 
a finite box.  Some of the conceptual difficulties with a possible stringy 
realization of de~Sitter space are connected to the fact that it is very 
difficult to confine strings in a finite box.  

\noindent There are also some qualitative differences between \godel\ and 
de~Sitter holography worth pointing out:  

\item{$\bullet$} In the \godel\ universe, the preferred screens are timelike, 
just as the canonical global screen in $AdS$ space.  On the other hand, the 
observer-dependent preferred screens in de~Sitter space are null.  

\item{$\bullet$} The \godel\ universe is supersymmetric.

\noindent 
In order to decide whether holography in the \godel\ universe can be 
used as a supersymmetric laboratory for exploring conceptual questions arising 
in de~Sitter holography (or more generally, holography in cosmological 
spacetimes), one needs a more microscopic understanding of the \godel\ 
universes in string and M-theory.  

\newsec{T-Duality of \godel\ Universes}

One can compactify one of the flat directions $\R^6$ (say $y_6$) of the 
M-theory \godel\ solution on $\S^1$ with constant radius $\CR$   
and obtain the following Type IIA \godel\ background, 
\eqn\eetypeiiag{\eqalign{ds^2&=-(dt+\beta\omega)^2+\sum_{i=1}^4(dx_i)^2+
\sum_{a=1}^5(dy_a)^2,\cr
H_{ij5}&=2\beta J_{ij},\cr
F_{ijab}&=2\beta J_{ij}K_{ab},\cr}}
where now in Type IIA theory $a,b\ldots=1,\ldots 5$. The dilaton is constant, 
implying that the string coupling $g_s$ can be kept small everywhere, and the 
\godel\ solution is a solution of weakly coupled Type IIA superstring theory.  
Now, we can T-dualize along various dimensions. 

\subsec{T-duality to a supersymmetric Type IIB pp-wave}

The $H$-field of the Type IIA \godel\ solution \eetypeiiag\ extends along 
$y_5$, the dimension that was paired up in M-theory with the extra dimension 
$y_6$. It turns our that T-duality along this dimension is particularly 
interesting. We first rename $y_5\equiv z$, and use the gauge in which 
\eqn\eebfield{B_{iz}=\beta J_{ij}x^j.}
Due to the absence of $g_{z\mu}$ cross-terms in the metric, no
$B$-field will be generated after T-duality, and one gets  
\eqn\eetdual{ds^2_{\rm IIB}=-dt^2-2\beta\omega\,(dt+dz)+\sum_{i=1}^4 dx_i^2
+\sum_{a=1}^4 dy_a^2+dz^2.}
To see that this Type IIB solution is in fact a supersymmetric pp-wave, it 
will be convenient to change the coordinates as follows.  First, define 
lightcone coordinates $u=t+z$, $v=t-z$, and also switch from the Cartesian 
coordinates $x_i$ to the ``bipolar'' coordinates given in \eefivebipol . Then, 
we perform a $u$-dependent rotation in each of the two preferred planes of 
rotation, 
\eqn\eeurot{\tilde\phi_i=\phi_i-\beta u.}
Upon introducing new Cartesian coordinates $\tilde x_i$, 
\eqn\eenewx{\eqalign{\tilde x_1+i\tilde x_2&=r_1e^{i\tilde\phi_1},\cr
\tilde x_3+i\tilde x_4&=r_2e^{i\tilde\phi_2},\cr}}
the Type IIB metric \eetdual\ T-dual to the \godel\ universe becomes 
\eqn\eetdualpp{ds^2_{\rm IIB}=-du\,dv-\beta^2(\sum_{i=1}^4 \tilde x_i^2)
du^2+\sum_{i=1}^4 d\tilde x_i^2+\sum_{a=1}^4 dy_a^2.}
This metric has the standard form of a supersymmetric pp-wave, with the 
\godel\ rotation parameter $\beta$ precisely equal to the conventionally 
normalized $\mu$ parameter of the pp-wave.    
One can also easily T-dualize the Ramond-Ramond fields:  The self-dual Type 
IIB five-form of the Type IIB solution is given by 
$F_5\sim du\wedge \tilde J\wedge K$, where 
$\tilde J=\sum_{i,j=1}^4J_{ij}d\tilde x_i\wedge d\tilde x_j$ and 
$K=\sum_{a,b=1}^4K_{ab}dy_a\wedge dy_b$. This Type IIB solution 
is in fact the supersymmetric pp-wave resulting from the Penrose limit of the 
near-horizon $AdS_3\times\S^3\times T^4$ geometry of a system of intersecting 
D3-branes, and was first found in \cvetic .  

\subsec{\godel/pp-wave T-duality}

We have shown that the Type IIA \godel\ universe is T-dual to a Type IIB 
pp-wave.  One can turn this observation around, and ask whether other known 
pp-waves can also be T-dual to new \godel-like universes.  We indeed find 
a rich picture of \godel/pp-wave duality which goes beyond the scope of the 
pp-wave T-dualities discussed in the literature (see, e.g., \tpp).  

Before generalizing the result of the previous subsection to a broader class 
of \godel/pp-wave pairs, it is instructive to first clarify which Killing 
dimension of the Type IIB pp-wave is being compactified on $\S^1$ and 
T-dualized to produce the Type IIA \godel\ universe.  Consider first the 
Killing vector 
\eqn\eekillbad{\xi_0=\frac{\p}{\p u}-\frac{\p}{\p v}}
of the Type IIB pp-wave background.  This vector is space-like at the origin, 
but becomes time-like at some critical radial distance.  One can remedy this 
problem by augmenting $\xi_0$ with a rotation in each of the two preferred 
planes 
\eqn\eekillgood{\xi=\frac{\p}{\p u}-\frac{\p}{\p v}
+\beta\left(\frac{\p}{\p\tilde\phi_1}+\frac{\p}{\p\tilde\phi_2}\right).}
This Killing vector $\xi$ is everywhere spacelike, with the space-like 
rotation off-setting the effect of the $du^2$ terms in the metric to keep this 
modified Killing vector spacelike.  Moreover, the norm of $\xi$ is 
\eqn\eenorm{|\xi|^2=1.}
Consequently, if we compactify the orbit of $\xi$ on a circle of fixed radius 
$R$ and T-dualize, the dilaton field of the resulting solution will stay 
constant.  This T-duality is precisely the inverse of the IIA $\rightarrow$ 
IIB T-duality that maps the \godel\ solution to the pp-wave.  
Note that closed timelike curves are introduced even though the orbifold 
action is generated by an everywhere-spacelike Killing vector. 

\subsec{New supersymmetric \godel\ universes in string and M-theory}

These observations lead to a very simple and general prescription for 
constructing a large class of \godel/pp-wave T-dual pairs.   Start with any 
pp-wave in which an analog of the Killing vector $\xi$ of \eekillgood\ (and 
satisfying \eenorm\ if we want constant $g_s$) can be identified.  
Compactification on $\S^1$ along this Killing direction followed by T-duality 
produces a \godel\ like solution of the T-dual string theory.  

As an example of this, we present a new supersymmetric \godel\ universe of 
Type IIA theory, as the T-dual of the maximally supersymmetric Type IIB 
pp-wave \iibpp .  Using the obvious generalization of \eekillgood\ that now 
involves four independent rotations in four independent two-planes of the 
pp-vave, we obtain a Type IIA geometry with a constant $H_3$ and $F_4$.  This 
Type IIA solution can be lifted to an M-theory solution of $\R^{11}$ 
topology.  Its metric factorizes to a product of a non-trivial metric on a 
$\CG_9$ factor and the flat metric on $\R^2$,  
\eqn\eegodelnine{\eqalign{ds^2&=-(dt+\beta\varpi)^2+\sum_{I=1}^8
(dx_I)^2+\sum_{A=1}^2(dy_A)^2,\cr
\varpi&=J_{IJ}x_Idx_J,\cr}}
and the four-form strength can be written as 
\eqn\eestrm{\eqalign{G_{ijkl}&=4\beta\epsilon_{ijkl},\cr
G_{mnpq}&=-4\beta\epsilon_{mnpq},\cr}\qquad
\eqalign{G_{ijAB}&=-2\beta J_{ij} K_{AB},\cr
G_{mnAB}&=-2\beta J_{mn} K_{AB},\cr}}
where $i,\ldots = 1, \ldots,4$ and $m,\ldots = 5, \ldots,8$, while the
indices $I,\ldots = 1, \ldots, 8$ and $A,B = 1,2$; all the non-zero
components of the K\"ahler forms $J_{IJ}$ and $K_{AB}$ are now given
by $J_{12}=-J_{21}=J_{34}=-J_{43}=J_{56}=-J_{65}=J_{78}=-J_{87}=1$ and
$K_{12}=-K_{21}=1$.

This new supersymmetric \godel\ solution $\CG_9\times\R^2$ of M-theory 
exhibits exactly the same qualitative holographic features as the 
$\CG_5\times\R^6$ solution.  In particular one finds compact closed timelike 
curves that are topologically large, and the analysis of geodesics reveals 
the same qualitative structure of holographic screens. 

\newsec{Discussion}

Following a phenomenological approach to holography, we have identified 
preferred holographic screens as seen by inertial observers in a class 
of homogeneous universes of the \godel\ type, with closed timelike curves.  
The structure of holographic screens change dramatically the question of 
causality, by hiding all closed timelike curves or breaking them into 
causal pieces.  It is tempting to suspect that holography serves as the 
chronology protection agency, and in combination with a version of the 
complementarity principle can lead to a consistent quantum mechanical 
description of this universe.  We also noticed close analogies with the 
structure of holographic screens in de~Sitter space, which can make the 
\godel\ universes an interesting supersymmetric laboratory for exploring 
de~Sitter holography.  This phenomenological identification of natural 
screens does not tell us, however, whether the holographic dual is given 
by some self-consistent quantum mechanics, or whether the pathology of 
closed timelike curves is just translated into some inconsistency of the 
holographic dual.  These and similar questions require a microscopic 
understanding of holography in \godel\ universes in string or M-theory.  
We have found evidence that the \godel-like cosmologies represent a 
remarkable and highly solvable class of solutions of string theory, and 
are in fact T-dual to solvable supersymmetric pp-wave solutions.  Further 
investigation of microscopic aspects of \godel\ universes and their 
holography in string and M-theory is in progress \elsewhere .  

\bigskip
\noindent{\bf Acknowledgement}
\smallskip
We wish to thank Raphael Bousso, Lisa Dyson, Ori Ganor, Eric Gimon, 
Mukund Rangamani, Harvey Reall, and Radu Tatar for useful discussions.  
This work was supported by the Berkeley Center for Theoretical Physics, by 
the NSF Grant PHY-0098840, and by DOE Grant DE-AC03-76SF00098.  This paper 
is an outcome of the bonus class project assigned by one of us (P.H.) in his 
string theory course at UC Berkeley in September 2002.  

\bigskip
\noindent{\bf Note Added}
\smallskip
As was pointed out to us after the completion of this work, the T-duality 
relation between the 5d \godel\ universe \eetypeiiag\ and the geometry of 
\eetdual\ can also be obtained from a T-duality relation 
found a week earlier by Herdeiro in \lastweek , by taking the limit of zero 
charges $P=Q=Q_{KK}=0$ in Eqn.~(4.11) of \lastweek .%
\foot{We thank Harvey Reall for pointing this out to us.}
However, it was not realized in \lastweek\ that the T-dual of the \godel\ 
universe is a supersymmetric Hpp-wave; instead, this T-dual was conjecturally 
interpreted in \lastweek\ as a rotating background.  

\appendix{A}{Geometry of the \godel\ Universes}

In this appendix we collect various aspects of the Riemannian geometry of 
the \godel\ universes $\CG_5$ and $\CG_9$ that play a central role 
in the paper.  

We are using the +++ conventions of MTW \mtw ; in particular, 
our metric is of the ``mostly plus'' signature.  

\bigskip
\noindent
{\it The five-dimensional \godel\ universe}
\smallskip

In the original Cartesian coordinates $t,x_i$ it is natural to introduce a 
vielbein 
\eqn\eefunf{e^0=dt+\beta\omega,\qquad e^i=dx_i,\quad i=1,\ldots 4,}
so that the metric on $\CG_5$ can be written simply as 
\eqn\eemetfunf{g_{\mu\nu}=-e_\mu{}^0e_\nu{}^0+\sum_i e_\mu{}^ie_\nu{}^i.}

In this vielbein, the spin connection one-forms are
\eqn\eespincon{\eqalign{\Omega_{ij}&=\beta J_{ij}dt+\beta^2 J_{ij}J_{k\ell}
x_k\,dx_\ell,\cr
\Omega_{i0}&=-\Omega_{0i}=\beta J_{ij}\,dx_j.\cr}}
These simple expressions for the spin connection can be used to easily 
extract the form of the Ricci tensor in the Cartesian coordinates, 
\eqn\eericci{R_{\mu\nu}dX^\mu dX^\nu=4\beta^2\,dt^2
+8\beta^3J_{ij}x_i\,dt\,dx_j+2\beta^2(\delta_{ij}
-2\beta^2J_{ik}J_{j\ell}x_kx_\ell)\,dx_i\,dx_j.}
The scalar curvature is constant, 
\eqn\eescalarcurv{R=4\beta^2,}
as is indeed implied by the homogeneity of the solution.  The  
Einstein tensor $G_{\mu\nu}=R_{\mu\nu}-\frac{1}{2}Rg_{\mu\nu}$ has 
the pressureless fluid form,  
\eqn\eeeinstein{\eqalign{G_{\mu\nu}dX^\mu dX^\nu&=6\beta^2dt^2+
12\beta^3J_{ij}x_i\,dx_j\,dt+6\beta^4J_{ik}J_{j\ell}x_kx_\ell\,dx_i\,dx_j\cr
&=6\beta^2u_\mu u_\nu dX^\mu dX^\nu,\cr}}
with 
\eqn\eeu{u_\mu dX^\mu=-dt-\beta J_{ij}x_i\,dx_j}
the covariant dual of the timelike Killing vector $\p/\p t$.  This is 
matched by the energy-momentum tensor of the constant gauge field strength 
$F\sim J$, which is also of the pressureless fluid form.  

{}For the calculation of the geodesic expansion parameter $\theta$ in the body 
of the paper, it is also useful to know the non-zero Christoffel symbols in 
the ``bipolar'' coordinates $(r_1,\phi_1,r_2\phi_2)$, 
\eqn\eechrist{\eqalign{\Gamma^{t}_{r_1t}&=\beta^2r_1,\vphantom{\frac{1}{2}}\cr
\Gamma^{t}_{\phi_1r_1}&=\beta^3r_1^3,\vphantom{\frac{1}{2}}\cr
\Gamma^{t}_{\phi_2r_1}&=\beta^3r_1r_2^2,\vphantom{\frac{1}{2}}\cr
\Gamma^{t}_{r_2t}&=\beta^2r_2,\vphantom{\frac{1}{2}}\cr
\Gamma^{t}_{\phi_2r_2}&=\beta^3r_2^3,\vphantom{\frac{1}{2}}\cr
\Gamma^{t}_{\phi_1r_2}&=\beta^3r_2r_1^2,\vphantom{\frac{1}{2}}\cr}
\qquad\eqalign{\Gamma^{r_1}_{\phi_1t}&=\beta r_1,\vphantom{\frac{1}{2}}\cr
\Gamma^{r_1}_{\phi_1\phi_1}&=-r_1(1-2\beta^2r_1^2),\vphantom{\frac{1}{2}}\cr
\Gamma^{r_1}_{\phi_1\phi_2}&=\beta^2r_1r_2^2,\vphantom{\frac{1}{2}}\cr
\Gamma^{r_2}_{\phi_2t}&=\beta r_2,\vphantom{\frac{1}{2}}\cr
\Gamma^{r_2}_{\phi_2\phi_2}&=-r_2(1-2\beta^2r_2^2),\vphantom{\frac{1}{2}}\cr
\Gamma^{r_2}_{\phi_1\phi_2}&=\beta^2r_1^2r_2,\vphantom{\frac{1}{2}}\cr}
\qquad\eqalign{\Gamma^{\phi_1}_{r_1t}&=-\frac{\beta}{r_1},\cr
\Gamma^{\phi_1}_{\phi_1r_1}&=\frac{1}{r_1}-\beta^2r_1,\cr
\Gamma^{\phi_1}_{\phi_2r_1}&=-\frac{\beta^2r_2^2}{r_1},\cr
\Gamma^{\phi_2}_{r_2t}&=-\frac{\beta}{r_2},\cr
\Gamma^{\phi_2}_{\phi_2r_2}&=\frac{1}{r_2}-\beta^2r_2,\cr
\Gamma^{\phi_2}_{\phi_1r_2}&=-\frac{\beta^2r_1^2}{r_2}.\cr}}

\bigskip
\noindent
{\it The nine-dimensional \godel\ universe}
\smallskip

This solution, discussed in Section~5, is T-dual to the maximally 
supersymmetric Type IIB pp-wave. 

We again introduce the natural vielbein in which the metric is of the form 
\eemetfunf ,
\eqn\eedevet{e^0=dt+\beta\omega,\qquad e^I=dx_I,\quad i=1,\ldots 8.}
In this basis, the spin connection one-forms are given by
\eqn\eespincon{\eqalign{\Omega_{IJ}&=\beta J_{IJ}dt+\beta^2 J_{IJ}J_{KL}
x_K\,dx_L,\cr
\Omega_{I0}&=-\Omega_{0I}=\beta J_{IJ}\,dx_J,\cr}}
with the Ricci tensor 
\eqn\eericcidevet{\eqalign{R_{MN}dX^MdX^N&=8\beta^2dt^2
+16\beta^3J_{IJ}x_I\,dx_J\,dt\cr
&\qquad{}+(2\beta^2\delta_{IJ}+8\beta^4J_{IK}x_KJ_{JL}x_L)
\,dx_I\,dx_J,\cr}}
the scalar curvature
\eqn\eescanine{R=8\beta^2,}
and the Einstein tensor
\eqn\eindevet{\eqalign{(R_{MN}-\frac{1}{2}Rg_{MN})dX^Mdx^N&=12\beta^2dt^2
+24\beta^3J_{IJ}x_I\,dx_J\,dt\cr
&\qquad{}-(2\beta^2\delta_{IJ}-12\beta^4J_{IK}x_KJ_{JL}x_L)\,
dx_I\,dx_J.\cr}}
Notice that unlike in the case of the five-dimensional \godel\ solution, the 
Einstein tensor of the nine-dimensional \godel\ universe is no longer of 
the pressureless fluid form.  
\listrefs
\end